\documentstyle[epsfig, aps,12pt,manuscript,amsbsy]{revtex}
\global\firstfigfalse \tighten \tightenlines

\begin{document}
\newcommand{\beq}{\begin{equation}}
\newcommand{\eeq}{\end{equation}}
\newcommand{\beqa}{\begin{eqnarray}}
\newcommand{\eeqa}{\end{eqnarray}}
\newcommand{\sr}{\sqrt}
\newcommand{\fr}{\frac}
\newcommand{\mn}{\mu \nu}
\newcommand{\G}{\Gamma}
\newcommand{\g}{ G_{n+2}}
\topmargin 0pt
\oddsidemargin 0mm

\begin{titlepage}
\begin{flushright}
INJE-TP-03-04 \\
hep-th/0304033
\end{flushright}

\vspace{5mm}
\begin{center}
{\Large \bf Brane-Bulk Interaction and Holographic Principle}

\vspace{12mm}

{\large
 Yun Soo Myung\footnote{email address:
 ysmyung@physics.inje.ac.kr}}
\vspace{8mm}

{ Relativity Research Center and School of Computer Aided Science \\ Inje University,
   Gimhae 621-749, Korea}
\vspace{8mm}

{\large
 Jin Young Kim\footnote{email address:
 jykim@kunsan.ac.kr}}

\vspace{8mm} { Department of Physics, Kunsan National University,
Kunsan 573-701, Korea \\ Asia Pacific Center for Theoretical
Physics, Pohang 790-784, Korea}

\end{center}
\vspace{5mm}
\centerline{{\bf{Abstract}}}
\vspace{5mm}

We introduce the brane-bulk interaction to discuss a limitation of
the cosmological Cardy-Verlinde formula  which is useful for  the
holographic description of brane cosmology. In the presence of the
brane-bulk interaction, we cannot find  the  entropy
representation of the first Friedmann equation (the cosmological
Cardy-Verlinde formula). In the absence of the interaction,  the
cosmological Cardy-Verlinde formula  is established even for  the
time-dependent charged AdS background. Hence, if there exists a
dynamic exchange of energy between the brane and the bulk (that
is, if $\tilde T^t~_y \not=0$), we cannot achieve  the
cosmological holographic principle on the brane.
\end{titlepage}

\newpage
\renewcommand{\thefootnote}{\arabic{footnote}}
\setcounter{footnote}{0} \setcounter{page}{2}

Recently the brane cosmology have received much
attention\cite{BDL1,BDL2,HCR,Krau,Ida} in connection with the
brane world scenario\cite{RS}. Further the brane cosmological
evolution was explored when the brane-bulk interaction
exists\cite{BDMP,HMR,DLSR,KKTTZ,Tet}. On the other hand the
cosmological holographic principle was realized  by applying  the
AdS/CFT correspondence~\cite{Verl,Lin,Noji,Wang,Brus,KPS,Youm}.

 In this
letter we  wish to clarify the relationship between the
cosmological holographic principle and  the brane-bulk energy
exchange.

We start with reviewing  the two important  issues.  One is a
static version called the Cardy-Verlinde formula which shows that
using the AdS/CFT correspondence~\cite{MAL},
 the entropy of a
conformal field theory (CFT) in any dimension can be expressed in
terms of a generalized form of the  Cardy formula~\cite{Card}.
Explicitly one can consider a CFT residing in an
$(n+1)$-dimensional spacetime with the  static metric for  the
Einstein space $ds^2_{CFT} =-d\tau^2 +r^2d\Omega_n^2$, where
$d\Omega^2_n$ denotes a unit $n$-dimensional sphere. For a
strongly coupled CFT with AdS dual, one can obtain the
Cardy-Verlinde formula which states that  an entropy ($S$) can be
expressed as a square root of the energy ($E$)
 \beq \label{1eq1} S=\frac{2\pi
r}{n}\sqrt{E_c(2E-E_c)} \eeq with the Casimir energy ($E_c$) for a
finite system. Indeed, this formula was checked to hold for
various kinds of AdS-bulk spacetimes; AdS-Schwarzschild black
hole~\cite{Verl}, AdS-Kerr black hole~\cite{Klem},
 AdS-charged black hole~\cite{Cai1}, AdS-Taub-Bolt
spacetimes~\cite{Birm1,Witten2}.

The other  is  a dynamic version called the  cosmological
Cardy-Verlinde formula which connects the Cardy-Verlinde formula
 Eq.(\ref{1eq1}) with the first Friedmann equation. We introduce the
Friedmann-Robertson-Walker (FRW) metric based on the isotropic and
homogeneous  universe
\begin{equation}
\label{1eq2} ds^2_{FRW} =-d\tau^2 +{\cal R}^2(\tau)d\Omega_n^2
\end{equation}
with a scale factor ${\cal R}(\tau)$. Although  two line elements
$dS_{CFT}^2$ and $dS_{FRW}^2$ have  different origins, these are
considered to be conformally equivalent. Hence it is possible to
make a connection between these if some conditions are satisfied.
 For a closed ($k=1$) radiation-dominated
 universe, two Friedmann equations are given by
\begin{eqnarray}
\label{1eq3}
&& H^2 =\frac{16 \pi G_{n+1}}{n(n-1)}\rho -\frac{1}{{\cal R}^2}, \\
\label{1eq4} && \dot{H}=-\frac{8 \pi G_{n+1}}{n-1}\left (\rho
+p\right) +\frac{1}{{\cal R}^2},
\end{eqnarray}
where $H=\dot{{\cal R}}/{\cal R}$ is the Hubble parameter and  the
dot stands for the differentiation with respect to the proper time
$\tau$. $\rho=E/V$ is the energy density where $E$ is the energy
of radiation-matter filling the universe, $V= {\cal R}^n V(S^n)$
is the volume of the universe,
 and $p=\rho/n$ denotes the radiation-pressure.  Verlinde\cite{Verl} have proposed that
the first Friedmann equation~(\ref{1eq3}) can be expressed in
terms of three cosmological holographic entropies
\cite{Beke,Bous}; Bekenstein-Verlinde entropy $ S_{
BV}=\frac{2\pi}{n}E {\cal R}$, Bekenstein-Hawking entropy $ S_{
BH}=(n-1)\frac{V}{4G_{n+1}{\cal R}}$,  Hubble  entropy $ S_{
H}=(n-1)\frac{HV}{4G_{n+1}}$. Then the first Friedmann
equation~(\ref{1eq3}) can  be rewritten as an entropy relation
\begin{equation}
\label{1eq5} S^2_{H}+ (S_{BV}-S_{BH})^2=S_{BV}^2.
\end{equation}
Making use of the Bekenstein-Verlinde entropy as
 $S_{BH} \equiv \frac{2\pi}{n} E_{ BH} {\cal R}$, we define the Bekenstein-Hawking energy
 $E_{BH}=\fr{n(n-1)V}{8 \pi G_{n+1} {\cal R}^2}$. Then
Eq.(\ref{1eq5}) leads to the cosmological Cardy-Verlinde formula
\begin{equation}
\label{1eq6} S_H=\frac{2\pi {\cal R}}{n}\sqrt{E_{ BH}(2E-E_{BH})}.
\end{equation}
 It
is straightforward to recover the Cardy-Verlinde formula of
Eq.(\ref{1eq1}) from Eq.(\ref{1eq6}) replacing $(S_H,E_{BH},{\cal
R})$ by ($S,E_c,r$) respectively. This apparent connection between
the static Cardy-Verlinde formula and the first Friedmann equation
comes mainly from the fact that two line elements ($ds^2_{CFT}$
and $ds^2_{FRW}$)  are conformally equivalent to each other. The
difference is their origin that one is static, while the other is
dynamic. This observation implies that the Friedmann equation and
the Cardy-Verlinde formula are closely related.  Furthermore it
was suggested that these can be derived from the same principle.

At this stage, we ask of how much  the Cardy-Verlinde formula of
Eq.(\ref{1eq1}) is useful for describing a cosmological
holographic principle in the brane cosmology \footnote{There exist
two approaches for the brane cosmology: one is the moving domain
wall (MDW) approach\cite{Krau,Ida} and  the other is
Binetruy-Deffayet-Langlois (BDL) approach\cite{BDL1,BDL2}.}. In
general, one says that if the dynamic equation of the brane (the
first Friedmann equation) is expressed in terms of the bulk
quantities, the cosmological holographic principle is realized.
 In
this direction, Savonije and Verlinde~\cite{Savo} have made the
first progress by introducing  a concrete model,
 an MDW in the ($n+2$)-dimensional  AdS-Schwarzschild black
hole. There exists a remarkable relationship between the
thermodynamics of a CFT  on the brane and the gravitational
dynamics of the brane in dual AdS space. It is well-known that the
dynamic equations governing the motion of the brane are exactly
given by the $(n+1)$-dimensional Friedmann equations. Actually all
moving domain walls in the AdS-Schwarzschild black hole may take a
kind of holographic matter $(\rho_h = E/V, E \sim m \ell/{\cal
R})$
 which comes originally
from the Schwarzschild  mass-term  $ m/r^{n-1}$. One identifies
this either a holographic term of strongly coupled CFT-radiation
matter filling the brane or a candidate for dark matter\cite{Col}.
In this work we keep a view of the holographic term. In this sense
one insists that the cosmological holography can be realized by
the extension of the AdS/CFT correspondence. Furthermore it was
shown that the cosmological Cardy-Verlinde formula Eq.(\ref{1eq6})
matches exactly  with the Cardy-Verlinde formula Eq.(\ref{1eq1})
 when the brane crosses the black hole horizon. This means that
 we can achieve the equalities $S=S_H,~E_c=E_{BH}$ at a holographic point $r=r_H={\cal
 R}_H$.
 This is an important result in the sense that from the Friedmann
 equations we know about thermodynamics of the CFT-radiation matter filled in the
 universe. The second Friedmann equation (\ref{1eq4}) leads to the
 Smarr-type
 formula of $ E_{BH}= n(E +pV-{\cal T}_H S_H)$ with the Hubble
 temperature defined by ${\cal T}_H=- \frac{\dot H}{2 \pi H}$.
 It is related to the defining equation of the Casimir energy on
 the CFT side: $E_c=n(E +pV -TS)$.

In the MDW approach the bulk spacetime is fixed. This means that
we can derive the gravitational dynamics of the brane  from the
given bulk black hole background. However, we don't know precisely
what kinds of  higher dimensional  theories we should start with,
even if string theories provide us some informations about it.
 In this sense the MDW approach is considered as a  restricted
 one  although it provides us an exact evolution of the brane.
On the other hand  the BDL approach is a genuine extension of  the
Kaluza-Klein cosmology to account for the local distribution on
the brane. In this case the location of the brane is fixed with
respect to the bulk direction.  Also this approach is useful for
describing the cosmological evolution of the brane when a
brane-bulk interaction exists. But there are some ambiguities in
interpreting the holographic matter. From now on, we follow the
BDL brane cosmology~\cite{BDL1,BDL2}.  We introduce the
Gaussian-normal bulk metric for ($1+3+1$)-dimensional spacetime
\begin{equation}
\label{1eq7} ds^2_{BDL} = -c^2(t,y) dt^2
+a^2(t,y)\gamma_{ij}dx^idx^j +b^2(t,y)dy^2,
\end{equation}
where $\gamma_{ij}$ is the metric of a three-dimensional space
with a constant curvature of $6k$.  Let us  express the bulk
Einstein equation $G_{ M N}=\frac{1}{2M^3} T_{ M N}$ in terms of
the BDL metric~\footnote{Our action is given by $S_5= \int d^5x
\sqrt{-g} \big(M^3R-\Lambda + \tilde {\cal L}^{mat}_B \Big) + \int
d^4x \sqrt{-\hat g}  {\cal L}^{mat}_b $ with $ M^3=1/16 \pi G_5,
~{\cal L}^{mat}_b=- \sigma +\cdots$. }. Now we introduce a
$(1+3)$-dimensional brane located at $y=0$. We choose for
simplicity the same stress-energy tensor $ T^M~_N={\rm
diag}(-\Lambda,-\Lambda,-\Lambda,-\Lambda,-\Lambda)+ \tilde T^M~_N
$ on both sides. Here the bulk cosmological constant is given by
$\Lambda$. The bulk stress-energy tensor $\tilde T^M~_N$ from
$\tilde {\cal L}^{mat}_B$ is not needed to have a specific form
initially. If $\tilde T^t~_y=0$, it is obvious that there is no
brane-bulk interaction in a dynamical way. The local stress-energy
tensor from ${\cal L}^{mat}_b$ including the brane tension
$\sigma$ is assumed to be of the form
\begin{equation}
\label{1eq8}  \tau^{\mu}~_\nu = \frac{\delta(y)}{b}{\rm
diag}(-\rho-\sigma,p-\sigma,p-\sigma,p-\sigma,0).
\end{equation}

We are interested in solving the Einstein equations at the
location of the brane. We indicate by the subscript ``0" the value
of the various quantities on the brane. Also it is convenient to
choose the Gaussian-normal gauge with $b_0=1$ and the temporal
gauge with $c_0=1$ on the brane. We obtain, from
$G_{0y}=\frac{1}{2M^3} T_{0y}$,
\begin{equation}
\label{1eq9} \dot \rho +3 \frac{\dot a_0}{a_0}\rho(1+\omega)= -2
\tilde T^0~_y.
\end{equation}
Here we assumed an equation of state $p=\omega \rho$. On the other
hand, the average part of $yy$-component equation is given by
\begin{equation}
\label{1eq10}
 \frac{\ddot a_0}{a_0} + \Big( \frac{\dot a_0}{a_0} \Big )^2 +
 \frac{k}{a_0^2} =\frac{1}{6M^3} \Big( \Lambda + \frac{\sigma^2}{12 M^3}
 \Big) -\frac{1}{144M^6}\Big(\sigma(3p -\rho)+\rho(3p+\rho)\Big)-
 \frac{1}{6M^3} \tilde T^y~_y.
  \end{equation}
 Then we rewrite
Eq.(\ref{1eq10}) in the following equivalent form
\begin{eqnarray}
\label{1eq11} && H^2_0 = \frac{1}{144M^6} \Big( \rho^2 + 2\sigma
\rho \Big) -\frac{k}{a^2_0} + \chi +\phi+ \frac{1}{12M^3} \Big(
\Lambda + \frac{\sigma^2}{12 M^3}
 \Big),  \\
 \label{1eq12}
 &&\dot \chi +4 H_0 \chi = \frac{1}{36M^6} (\rho+\sigma) \tilde
 T^0~_y,\\
  \label{1eq13}
 && \dot \phi +4 H_0 \phi =- \frac{1}{3M^3} H_0 \tilde T^y~_y ,
\end{eqnarray}
with $H_0=\dot a_0/a_0$. In the case of $p=\rho=0$ and
$\phi=\chi=0$, we wish to find the Randall-Sundrum
vacuum\cite{RS}. Hence we choose the bulk cosmological constant
$\Lambda = - \sigma^2 / 12M^3 = -12M^3 / \ell^2$ and the brane
tension $\sigma=12 M^3 /\ell$ to have a critical brane. Hence the
cosmological evolution is determined by four initial parameters
$(\rho_i,a_{0i},\chi_i,\phi_i)$ instead of the two
$(\rho_i,a_{0i})$. This is so because the generalized Friedmann
equation (\ref{1eq11}) is not a first integral of the Einstein
equations. It is mainly due to the energy exchange $\tilde T^t~_y
$ between the brane and the bulk. In the case of $\phi=0,\tilde
T^t~_y=A \rho>0$ and negligible $\tilde T^y~_y$, one finds a
mirage-radiation term $\chi \sim (1-e^{-At/2})/a_{0}^4$ for an
energy outflow from the brane\cite{KKTTZ}. It can be considered
that the real brane matter decays into the extra dimension. Also
for $\phi=0, \tilde T^t~_y \sim -\frac{1}{a_0^q}$ and negligible
$\tilde T^y~_y$, it is shown that the energy influx from the bulk
generates a cosmological acceleration on the brane with the
acceleration parameter $Q \equiv\frac{1}{H_0^2} \frac{\ddot
a_0}{a_0}=1-\frac{q}{3}$ where $0 \le q \le 4$ \cite{Tet}.
However, in general it will be a formidable task to obtain the
solution of Eqs.(\ref{1eq9})-(\ref{1eq13}) because it gives rise
to a complicated dynamics between the brane and the bulk.
Importantly it is very hard to express Eq.(\ref{1eq11}) in terms
of the cosmological Cardy-Verlinde formula of Eq.(\ref{1eq6})
because Eq.(\ref{1eq11}) is not a first integral of the Einstein
equations.

 In this work we are interested in exploring the relationship between
the brane-bulk interaction and the (cosmological) holographic
principle. In the case of $\tilde T^t~_y=\tilde T^y~_y=0$, it is
shown  from Eqs.(\ref{1eq12})-(\ref{1eq13})  that one finds the
mirage or Weyl-radiation terms of $\chi =C_{\chi}/a_0^4$ and $\phi
=C_{\phi}/a_0^4$ with two unknown constants $C_{\chi},C_{\phi}$.
In view of the moving domain wall approach these correspond to the
holographic-radiation terms originated from the mass of the
AdS-black hole\cite{Kim}. Furthermore, it is obvious that the
omission of $\phi$  is not justified when the energy influx
(outflux) stops and thus $\tilde T^t~_y$ becomes zero. The authors
in\cite{KKTTZ,Tet} assumed that at this moment, $\tilde T^y~_y$ is
also very small, so that $\phi$ gives a negligible contribution to
Eq.(\ref{1eq11}). However, this is not the general case to obtain
the information on the brane.

We propose here  that {\it as far as there exists any dynamical
interaction between the brane and the bulk, one cannot establish
the cosmological holographic principle}. The reason is as follows.
First, up to now one does not  find  any dynamical localization
mechanism for the brane matter by choosing an appropriate
interaction term $\tilde T^t~_y$. This means that we do not
specify the details of the mechanism through which energy is
transferred from the bulk to the brane. Second the holographic
principle is based on the static non-local mechanism, for example,
the projection of the bulk Weyl tensor  onto the brane\cite{SMS}.
The AdS/CFT correspondence is a tool to realize the holographic
principle. In the case of brane cosmology, we obtain the
holographic energy density $\rho_h$ on the brane by the extension
of  the AdS/CFT correspondence. Actually, one does not want to
have an environment where a dynamical energy exchange between the
brane and bulk exists. This is so because if $\tilde T^t~_y
\not=0$, one cannot identify which part belongs to the brane and
which part belongs to the bulk.  Finally, if $\tilde T^t~_y =0$,
Eq.(\ref{1eq11}) becomes a first integral (that is, the first
Friedmann equation). As a result, to establish the holographic
principle, one has to meet the condition of no dynamical exchange
between the brane and bulk : $\tilde T^t~_y =0$.

In order to see this with a specific example, let us introduce the
bulk Maxwell field whose solution takes the form under the metric
of Eq.(\ref{1eq7}) \cite{CMO,Myung2}
\begin{equation}
\label{1eq14} F_{yt} =\frac{{\cal Q}bc}{a^n},
\end{equation}
where ${\cal Q}$ is an unknown integration constant which is
proportional to the charge of the AdS-charged black hole if one
makes a further connection to the MDW approach. Then the bulk
stress-energy tensor $\tilde T^M~_N=4M^3(F^{MP}F_{NP}-\delta^M~_N
F^2/4)$ is given by\footnote{To obtain this, one takes $\tilde
{\cal L}^{mat}_B= -M^3F^2$ in footnote 2.}
\begin{equation}
\label{1eq15} \tilde T^M~_N =2M^3 {\rm diag}\left (
  -\frac{{\cal Q}^2}{a^{6}},  +\frac{{\cal Q}^2}{a^{6}},
  \cdots ,-\frac{{\cal Q}^2}{a^{6}} \right).
\end{equation}
This implies that on the brane, $\tilde T^t~_y=0, \tilde
T^y~_y=-2M^3{\cal Q}^2/a_0^6$. Although this  is a time-dependent
bulk term, there
 does not exist any dynamical exchange between the brane and bulk. Substituting these into
Eqs.(\ref{1eq12})-(\ref{1eq13}), one finds that
\begin{equation}
\label{1eq16}
 \phi=-{\cal Q}^2/3a^6_0,~~~~\chi={\cal
C}_{\chi}/a^4_0.
\end{equation}
 Equation (\ref{1eq11}) leads to the first Friedmann equation
 \begin{equation}
 \label{1eq17}
 H_0^2 = -\frac{k}{a_0^2} +\frac{{\cal C}_{\chi}}{a_0^4} -\frac{1}{3}
\frac{{\cal Q}^2}{a_0^6}  + \frac{1}{144M^6} \Big( \rho^2 +
2\sigma \rho \Big).
 \end{equation}
  This is an equation which is derived from the BDL approach and governs
the evolution of the fixed  brane  sandwiched in the  charged AdS
bulks. For $\rho= p=0$ case, let us  compare our equation
(\ref{1eq17}) with that of the moving domain wall in the charged
topological-AdS (CTAdS) black holes  which can be rewritten
as\cite{BM,Myung1,KT}
\begin{equation}
\label{1eq18} H^2 =-\frac{k}{{\cal R}^2} +\frac{\omega_3 M}{{\cal
R}^4} -\frac{3 \omega^2_3 Q^2}{16{\cal R}^6},
\end{equation}
where $\omega_3=1/3M^3V(S^3)$. This can be written as the
cosmological Cardy-Verlinde formula as
\begin{equation}
\label{1eq19} S_H=\frac{2\pi {\cal R}}{3}\sqrt{E_{ BH}[2(E-\Phi
Q/2)-kE_{BH}]},
\end{equation}
with the electric potential $\Phi=3\ell\omega_3Q/16{\cal R}^3$ on
the brane. Further its time-derivative is given by
\begin{equation}
\label{1eq20} kE_{BH}= 3(E +pV-\Phi Q-{\cal T}_H S_H).
\end{equation}
With the identification,
\begin{equation}
\label{1eq21} a_0 \leftrightarrow {\cal R},~~ {\cal C}_{\chi}
\leftrightarrow \omega_3 M,~~~ {\cal Q} \leftrightarrow \pm
\frac{3 \omega_3 Q}{4}
\end{equation}
our equation (\ref{1eq17}) with $p=\rho=0$  coincides with the
cosmological Cardy-Verlinde formula Eq.(\ref{1eq19}) in the
presence of the background charge. This shows that the
cosmological holographic principle is achieved.

 In case of Eq.~(\ref{1eq18}) for
the moving brane, the bulk is just fixed as the  two CTAdS black
holes with the mass $M$ and electric charge $Q$. These two
parameters encode  information of the bulk and are used to
describe a strongly coupled CFT  on the brane via the extension of
the AdS/CFT correspondence. On the other hand, in deriving
Eq.~(\ref{1eq17}), we don't need to know
 about the precise form of the AdS bulk geometry but we have to keep up
a bulk Maxwell field.  However, the two integration constants
${\cal C}_{\chi}$ and ${\cal Q}$ certainly encode  information of
the bulk, as the initial conditions. Just as in the moving domain
wall approach in which the brane moves in the fixed bulk and acts
as the boundary of the bulk, the Friedmann equation of the BDL
brane sandwiched in two AdS bulks carries the information of the
bulk as the initial conditions. This  shows that it is possible to
achieve  the cosmological holographic principle in the BDL
approach. Furthermore, to make a further connection with the
static Cardy-Verlinde formula, we have to choose the CTAdS black
hole background as the fixed bulk\cite{CMO,Myung2}.

In conclusion, if there exists a dynamic exchange of energy
between the brane and the bulk (that is, if $\tilde T^t~_y
\not=0$), one cannot achieve  the cosmological holographic
principle on the brane.

\section*{Acknowledgements}

This work was supported in part by  KOSEF, Project Number
 R02-2002-000-00028-0.



\begin{thebibliography}{99}
\bibitem{BDL1}P. Binetruy, C. Deffayet and D. Langlois, Nucl. Phys.
     {\bf B565}, 269 (2000) [hep-th/9905012].
\bibitem{BDL2}P. Binetruy, C. Deffayet, U. Ellwanger and D. Langlois,
      Phys. Lett. {\bf B477}, 285 (2000) [hep-th/9910219].
\bibitem{HCR} H. Chamblin and H. Real, Nucl. Phys. B {\bf 562},133 (1999)
          [hep-th/9903225].
\bibitem{Krau}P. Kraus, JHEP {\bf 9912}, 011 (1999) [hep-th/9910149].
\bibitem{Ida}D. Ida, JHEP {\bf 0009}, 014 (2000) [gr-qc/9912002].
\bibitem{RS} L. Randall and R. Sundrum, Phys. Rev. Lett. {\bf 83}, 4690 (1999)
              [hep-th/9906064].
\bibitem{BDMP} C. van de Bruck, M. Dorca, C. J. A. P. Martins, and M. Parry,
               Phys. Lett. B {\bf 495}, 183 (2000) [hep-th/0009056].
\bibitem{HMR} A. Hebecker and J. March-Russell,
      Nucl. Phys. B  {\bf 608}, 375(2001) [hep-th/0103214].
\bibitem{DLSR} D. Langlios, L. Sorbo and M. Rodriguez-Martinez,
      Phys. Rev. Lett. {\bf 89}, 171301(2002) [hep-th/0206146].
\bibitem{KKTTZ} E. Kiritsis, G. Kofinas, N. Tetradis, T. N. Tomaras
      and V. Zarikas, JHEP {\bf 0302},035(2003) [hep-th/0207060].
\bibitem{Tet} N. Tetradis, ``Cosmological Acceleration from Energy Influx",  hep-th/0211200.
\bibitem{Verl}E. Verlinde, ``On the Holographic Principle in a Radiation Dominated Universe", hep-th/0008140.
\bibitem{Lin} F.-L. Lin, Phys. Rev. D {\bf 63},  064026
               (2001) [hep-th/0010127].
\bibitem{Noji}S. Nojiri and S. Odintsov, Int. J. Mod. Phys. A {\bf 16},3273 (2001)
             [hep-th/0011115]; Class. Quant. Grav. {\bf 18}, 5227(2001)
              [hep-th/0103078].
\bibitem{Wang} B. Wang, E. Abdalla and R.K. Su,  Phys. Lett. B {\bf 503}, 394(2001)
              [hep-th/0101073].
\bibitem{Brus}R. Brustein, S. Foffa and G. Veneziano, Phys. Lett. B {\bf 507}, 270 (2001)
             [hep-th/0101083].
\bibitem{KPS}D. Klemm, A.C. Petkou, G. Siopsis and D. Zanon, Nucl. Phys. B {\bf 620}, 519 (2002)
                [hep-th/0104141].
\bibitem{Youm}D. Youm, Mod. Phys. Lett. A {\bf 16}, 1236 (2001) [hep-th/0105036].
\bibitem{MAL} J. Maldacena, Adv. Theor. Math. Phys. {\bf 2}, 231 (1998)
              [hep-th/9711200]; S. S. Gubser, I. Klebanov, A. Polyakov, Phys. Lett. B {\bf 428}, 105 (1998)
               [hep-th/9802109]; E. Witten, Adv. Theor. Math. Phys. {\bf 2}, 253 (1998)
                [hep-th/9802150].
\bibitem{Card}J.L. Cardy, Nucl. Phys. {\bf B270}, 186 (1986).
\bibitem{Klem}D. Klemm, A.C. Petkou and G. Siopsis, Nucl. Phys. B {\bf 601}, 380 (2001)
                 [hep-th/0101076].
\bibitem{Cai1} R.G. Cai, Phys. Rev. D {\bf 63}, 124018 (2001) [hep-th/0102113].
\bibitem{Birm1}D. Birmingham and S. Mokhtari, Phys. Lett. B {\bf 508}, 365 (2001)
                 [hep-th/0103108].
\bibitem{Witten2}E. Witten, Adv. Theor. Math. Phys. {\bf 2}, 505 (1998)
     [hep-th/9803131].

\bibitem{Beke}J.D. Bekenstein, Phys. Rev. D {\bf 23}, 287 (1981).
\bibitem{Bous}R. Bousso, JHEP {\bf 0004}, 035 (2001) [hep-th/0012052];
    JHEP {\bf 0011}, 038 (2000) [hep-th/0010252].

\bibitem{Savo}I. Savonije and E. Verlinde, Phys. Lett. B {\bf 507}, 305 (2001)
                [hep-th/0102042].
\bibitem{Col} A. Coley, Phys. Rev. D {\bf 66}, 023512 (2002)
                [hep-th/0110049].
\bibitem{Kim} N.J. Kim, H.W. Lee, Y.S. Myung and G. Kang, Phys. Rev. D {\bf 64}, 064022 (2001)
               [hep-th/0104159].
\bibitem{SMS} T. Shiromizu, K. Maeda, and M. Sasaki,  Phys. Rev. D {\bf 62},
            024012 (2000) [gr-qc/9910076].
\bibitem{CMO} R.G. Cai, Y. S. Myung and N. Ohta, Class. Quant. Grav. {\bf 18}, 5429 (2001)
              [hep-th/0105070].
\bibitem{Myung2} Y. S. Myung,  Class. Quant. Grav. {\bf 20}, 935 (2003) [hep-th/0208086].
\bibitem{BM} A.K. Biswas and S. Mukherji, JHEP {\bf 0103}, 046 (2001)
    [hep-th/0102138].
\bibitem{Myung1} Y. S. Myung, Mod. Phys. Lett. A {\bf 17},
                        1915(2002) [hep-th/0103241].
\bibitem{KT} P. Kanti and K. Tamvakis, ``Challenges and Obstacles for a Bouncing Universe in Brane Models",
 hep-th/0303073.

\end{thebibliography}
\end{document}